# Scientific Workflow Repeatability through Cloud-Aware Provenance

Khawar Hasham, Kamran Munir, Jetendr Shamdasani, Richard McClatchey
Centre for Complex Computing Systems (CCCS)
University of the West of England, UWE,
Bristol, United Kingdom
khawar.ahmad@cern.ch

*Abstract*—The transformations, analyses and interpretations of data in scientific workflows are vital for the repeatability and reliability of scientific workflows. This provenance of scientific workflows has been effectively carried out in Grid based scientific workflow systems. However, recent adoption of Cloud-based scientific workflows present an opportunity to investigate the suitability of existing approaches or propose new approaches to collect provenance information from the Cloud and to utilize it for workflow repeatability in the Cloud infrastructure. The dynamic nature of the Cloud in comparison to the Grid makes it difficult because resources are provisioned on-demand unlike the Grid. This paper presents a novel approach that can assist in mitigating this challenge. This approach can collect Cloud infrastructure information along with workflow provenance and can establish a mapping between them. This mapping is later used to re-provision resources on the Cloud. The repeatability of the workflow execution is performed by: (a) capturing the Cloud infrastructure information (virtual machine configuration) along with the workflow provenance, and (b) re-provisioning the similar resources on the Cloud and re-executing the workflow on them. The evaluation of an initial prototype suggests that the proposed approach is feasible and can be investigated further.

## I. INTRODUCTION

The scientific community is experiencing a data deluge due to the generation of large amounts of data in modern scientific experiments that include projects such as DNA analysis [1], the Laser Interferometer Gravitational Wave Observatory (LIGO) [2], the Large Hadron Collider (LHC) [3], and projects such as neuGRID [4] and neuGRIDforUser [43, 44]. In particular the neuGRID community is utilising scientific workflows to orchestrate the complex processing of its data analysis. A large pool of compute and data resources are required to process this data, which has been available through the Grid [5] and is now also being offered by the Cloud-based infrastructures.

Cloud computing [6] has emerged as a new computing and storage paradigm, which is dynamically scalable and usually works on a pay-as-you-go cost model. It aims to share resources to store data and to host services transparently among users at a massive scale [7]. Its ability to provide an on-demand computing infrastructure with scalability enables distributed processing of complex scientific workflows [8] for the scientific community. Recent work [29] is now experimenting with Cloud infrastructures to assess the feasibility of executing workflows on the Cloud.

An important consideration during this data processing is to gather data that can provide detailed information about both the input and the processed output data, and the processes involved to verify and repeat a workflow execution. Such a data is termed as "Provenance" in the scientific literature. Provenance is defined as the derivation history of an object [9]. This information can be used to debug and verify the execution of a workflow, to aid in error tracking and reproducibility. This is of vital importance for scientists in order to make their experiments verifiable and repeatable. This enables them to iterate on the scientific method, to evaluate the process and results of other experiments and to share their own experiments with other scientists [10]. The execution of scientific workflows in Clouds brings to the fore the need to collect provenance information, which is necessary to ensure the reproducibility of these experiments on Cloud [11].

The dynamic and geographically distributed nature of Cloud computing makes the capturing and processing of provenance information a major research challenge [12, 13, 14, 29]. Contrary to Grid computing, the resources in the Cloud computing are virtualised and provisioned *on-demand*, and released when a task is complete [1]. Generally, an execution in Cloud based environments occurs transparently to the scientist, i.e. the Cloud infrastructure behaves like a black box. Therefore, it is critical for scientists to know the parameters that have been used, what execution environment was used, and what data products were generated in each execution of a given workflow [15]. Due to the dynamic nature of the Cloud the exact resource configuration should be known in order to reproduce the execution environment. Due to these reasons, there is a need to capture information about the Cloud infrastructure along with workflow provenance, to aid in the repeatability of experiments. There has been a lot of research related to provenance in the Grid [16, 17, and 18] and a few initiatives [19, 20, 21] for the Cloud. However, they lack the information that can be utilized for re-provisioning of resources on the Cloud, thus they cannot create the similar execution environment(s) for workflow repeatability. This paper presents a theoretical description of an approach that can augment workflow provenance with infrastructure level information of the Cloud and use it to establish similar execution environment(s) and repeat a given workflow. Important points discussed in this paper are as follows: section II presents the motivation to capture Cloud infrastructure information. In section III, a workflow execution in a Cloud scenario is discussed. Section IV presents an overview of the proposed approach. Section V presents a basic evaluation of the developed prototype. Section VI presents some related work. And finally section VII presents some conclusions and directions for future work.

## II. MOTIVATION

A research study [42] conducted to evaluate the reproducibility of scientific workflows has shown that around 80% of the workflows can not be reproduced, and 12% of them are due to the lack of information about the execution environment. This becomes more challenging issue in the context of Cloud. As discussed above, the Cloud presents a dynamic environment in which resources are provisioned on-demand. For this, a user submits resource configuration information as resource provision request to the Cloud infrastructure. A resource is allocated to the user if the Cloud infrastructure can meet the submitted resource configuration requirements. Moreover, the pay-as-you-go model in the Cloud puts constraints on the lifetime of a Cloud resource. For instance, one can acquire a resource for a lifetime but he has to pay for that much time. This means that a resource is released once a task is finished or payment has ceased. In order to acquire the same resource, one needs to know the configuration of that old resource. This is exactly the situation with repeating a workflow experiment on the Cloud. In order to repeat a workflow execution, a researcher should know the resource configurations used earlier in the Cloud. This enables him to re-provision similar resources and repeat workflow execution.

As the cost model for Cloud resources decreases with multiple providers in the market, it is possible to think of this factor as negligible. However, such provenance information is necessary especially for experiments in which performance is an important factor. For instance, a data-intensive job can perform better with 2 GB of RAM because it can accommodate more data in RAM which is a faster than hard disk. However, its performance will degrade if a resource of 1 GB RAM is allocated to this job as less data can be placed in RAM. Therefore, it is important to collect the Cloud infrastructure or virtualization layer information along with the workflow provenance to recreate similar execution environment to ensure workflow repeatability. In this paper, the terms "Cloud infrastructure" and "virtualization layer" are used interchangeably.

## III. A WORKFLOW EXECUTION SCENARIO ON THE CLOUD

In this section, a scenario (Fig. 1) is presented that can be used to execute a workflow on the Cloud. This has been discussed and tested by Groth et al. [22]. It uses Pegasus [23] as a workflow management system (WMS) along with the Condor [24] cluster on the Cloud infrastructure to execute workflow jobs.

A scientist creates his own workflow using a workflow authoring tool or uses an existing workflow from the *Workflow Provenance Store* e.g. Pegasus database and submits it to the Cloud infrastructure through Pegasus. Pegasus interacts with a cluster of compute resources in the form of Condor instances running on the virtual machines (VM) in the Cloud. Each VM has a Condor instance to execute the user's job.

Pegasus schedules the workflow jobs to these Condor instances and retrieves the workflow provenance information supported by the Pegasus database. The collected provenance information, which is stored in the Pegasus database, comprises of job arguments (input and outputs), job logs (output and error) and host information. However, the collected host information is not sufficient to re-provision resources on the Cloud because Pegasus was designed initially for the Grid environment, and it lacks this capability at the moment.

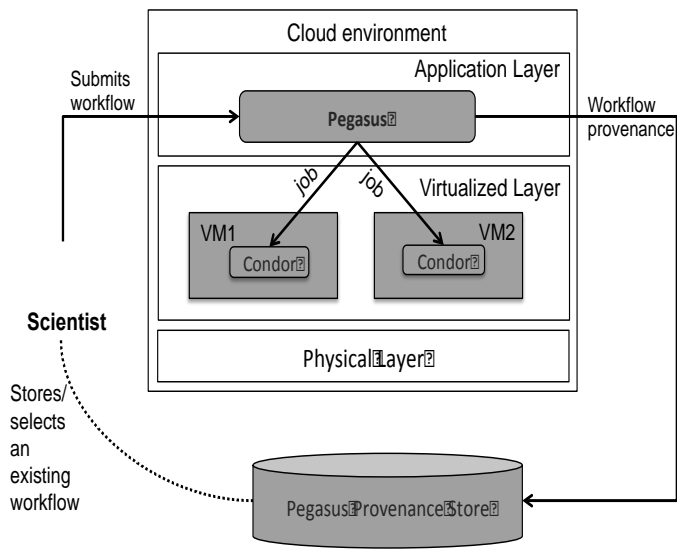

Fig. 1. Workflow execution setup on the Cloud infrastructure

## IV. PROPOSED APPROACH

An abstract view of the proposed architecture is presented in this section. This architecture is designed after evaluating the existing literature and keeping in mind the objectives of this research study. The proposed architecture is inspired by the mechanism used by Groth et al. [22] (as discussed in section III) for executing workflows on the Cloud. Fig. 2 illustrates the proposed architecture that can be used to capture the Cloud infrastructure information and to interlink it with the workflow provenance collected from a workflow management system such as Pegasus. This augmented or extended provenance information compromising of workflow provenance and the Cloud infrastructure information is named as Cloud-aware provenance (Fig. 2). The components of this architecture are briefly explained below.

- **Workflow Provenance:** This component is responsible for receiving provenance captured at the application level by the workflow management system (Pegasus). Since workflow management systems may vary, a plugin-based approach is used for this component. Common interfaces are designed to develop plugins for different workflow management systems. The plugin will then translate the workflow provenance according to the representation that will be used to interlink the workflow provenance along with other provenance information coming from the Cloud infrastructure.

- **Cloud Layer Provenance**: This component is responsible for capturing information collected from different layers of the Cloud. To achieve re-provisioning of resources on Cloud, this component focuses on the virtualization layer and retrieves information related to the Cloud infrastructure i.e. virtual machine configuration. This component is discussed in detail in section IV (A).
- **Provenance Aggregator:** This is the main component tasked to collect and interlink the provenance coming from different layers as shown in Fig. 2. It establishes interlinking connections between the workflow provenance and the Cloud infrastructure information. The provenance information is then represented in a single format that could be stored in the provenance store through the interfaces exposed by the Provenance API.
- **Workflow Provenance Store**: This data store is designed to store workflows and their associated provenance. This also keeps mapping between workflow jobs and the virtual compute resources they were executed on in the Cloud infrastructure.
- **Provenance API**: This acts as a thin layer to expose the provenance storage capabilities to other components. Through its exposed interfaces, outside entities such as the Provenance Aggregator would interact with it to store the workflow provenance information. This approach gives flexibility to implement authentication or authorization in accessing the provenance store.

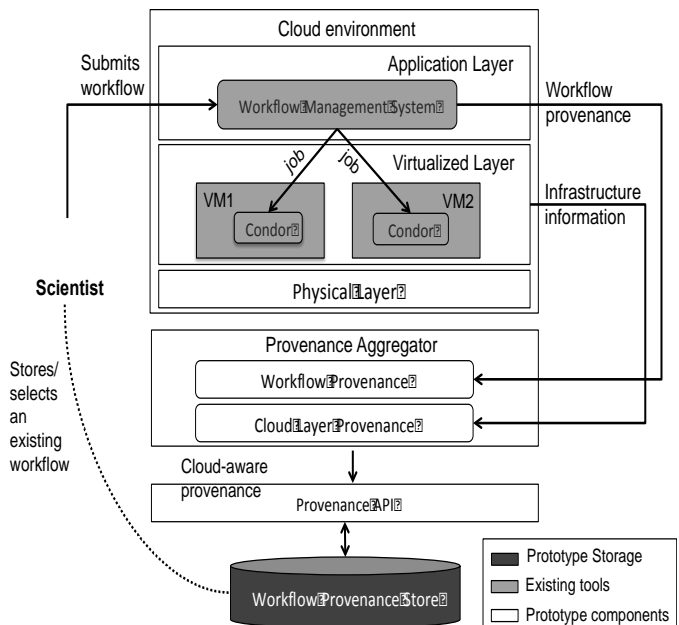

Fig. 2. An abstract architecture of the proposed approach

### A. Cloud Infrastructure Information Collection

The Cloud Layer Provenance component is designed in a way that interacts with the Cloud infrastructure to obtain the resource configuration information. As mentioned earlier, this information is later used for reprovisioning the resources to provide a similar execution infrastructure to repeat a workflow execution. Once a workflow is executed, Pegasus collects the provenance and stores it in its own internal database. Pegasus also stores the IP address of the virtual machine (VM) where the job is executed. However, it lacks other VM specifications such as RAM, CPUs, hard disk etc. The Cloud Layer Provenance component retrieves all the jobs of a workflow and their associated VM IP addresses from the Pegasus database. It then collects a list of virtual machines owned by a respective user from the Cloud middleware. Using the IP address, it establishes a mapping between the job and the resource configuration of the virtual machine used to execute the job. The resource configuration includes information about the flavour and image. *flavour* defines resource configuration such as RAM, Hard disk and CPUs, and *image* defines the operating system image used in that particular resource. By combining these two parameters together, one can provision a resource on the Cloud infrastructure. This information i.e. Cloud-aware provenance is then stored in the Provenance Store.

### B. Repeat Workflow using Cloud-Aware Provenance

In section IV (A), the job to Cloud resource mapping using provenance information has been discussed. This mapping is stored in the database for workflow repeatability purposes. In order to repeat a workflow, researcher first needs to provide the *wfID* (workflow ID), which is assigned to every workflow in Pegasus, to the proposed framework to re-execute the given workflow using the Cloud-aware provenance. It retrieves the given workflow from the Provenance Store database along with the Cloud resource mapping stored against this workflow. Using this mapping information, it retrieves the resource flavour and image configurations, and provisions the resources on Cloud. Once resources are provisioned, it submits the workflow. At this stage, a new workflow ID is assigned to this newly submitted workflow. This new wfID is passed over to the Provenance Aggregator component to monitor the execution of the workflow and start collecting its Cloud-aware provenance information. This is important to recollect the provenance of the repeated workflow, as it will enable us to verify the provisioned resources by comparing their resource configurations with the old resource configuration.

## V. PROTOTYPE EVALUATION

A Python based prototype has been developed using Apache Libcloud[1], a library to interact with the Cloud middleware. The presented evaluation of the prototype is very basic currently, however, as this work progresses further a full evaluation will be conducted. To evaluate this prototype, a 20 core Cloud infrastructure is acquired from Open Science Data Cloud (OSDC) organisation[2]. This Cloud infrastructure uses OpenStack middleware[3] to provide the infrastructure-as-a-Service capability. A small Condor cluster of three virtual machines is also configured. In this cluster, one machine is a master node, which is used to submit workflows, and the remaining two are compute nodes. These compute nodes are used to execute workflow jobs.

---

[1] Apache Libcloud – http://libcloud.apache.org [Last access 05-08-2014]
[2] OSDC- https://www.opensciencedatacloud.org/ [Last access 05-08-2014]
[3] OpenStack middleware – http://openstack.org [Last access 05-08-2014]

Table I. Cloud infrastructure mapped to the jobs of workflow with ID 114

| WF ID | Host IP | nodename | Flavour Id | minRAM (MB) | minHD (GB) | vCPU | Image name | Image id |
|---|---|---|---|---|---|---|---|---|
| 114 | 172.16.1.49 | osdc-vm3.novalocal | 2 | 2048 | 20 | 1 | wf_peg_repeat | f102960c- 557c-4253-8277-2df5ffe3c169 |
| 114 | 172.16.1.98 | mynode.novalocal | 2 | 2048 | 20 | 1 | wf_peg_repeat | 102960c- 557c-4253-8277-2df5ffe3c169 |

Table II. Cloud infrastructure information of repeated workflows (WfIDs117 and 122) after repeating the workflow 114

| WF ID | Host IP | nodename | Flavour Id | minRAM (MB) | minHD (GB) | vCPU | Image name | Image id |
|---|---|---|---|---|---|---|---|---|
| 117 | 172.16.1.183 | osdc-vm3-rep.novalocal | 2 | 2048 | 20 | 1 | wf_peg_repeat | f102960c- 557c-4253-8277-2df5ffe3c169 |
| 117 | 172.16.1.187 | mynode-rep.novalocal | 2 | 2048 | 20 | 1 | wf_peg_repeat | f102960c- 557c-4253-8277-2df5ffe3c169 |
| 122 | 172.16.1.114 | osdc-vm3-rep.novalocal | 2 | 2048 | 20 | 1 | wf_peg_repeat | f102960c- 557c-4253-8277-2df5ffe3c169 |
| 122 | 172.16.1.112 | mynode-rep.novalocal | 2 | 2048 | 20 | 1 | wf_peg_repeat | f102960c- 557c-4253-8277-2df5ffe3c169 |

Using the Pegasus APIs, a basic wordcount workflow application composed of four jobs is written. This workflow has both control and data dependencies [32] among its jobs, which is a common characteristic in scientific workflows. The first job (*Split* job) takes a text file and splits it into two files of almost equal length. Later, two jobs (*Analysis* jobs), each of these takes one file as input, then calculates the number of words in the given file. The fourth job (*merge* job) takes the outputs of earlier analysis jobs and calculates the final result i.e. total number of words in both files.

This workflow is submitted using Pegasus. The WfID assigned to this workflow is 114. The collected Cloud resource information is stored in database. Table I. shows the provenance mapping records in the Provenance Store for this workflow. The collected information includes the flavour and image (image name and Image id) configuration parameters. The Image id uniquely identifies an OS image hosted on the Cloud and this image contains all the software or libraries used during job execution. As an image defines all the required libraries of a job, the initial prototype does not extract the installed libraries information from the virtual machine at the moment. However, this can be done in future iterations to enable the proposed approach to reconfigure a resource at runtime on the Cloud.

The repeatability of the workflow achieved using the proposed approach (discussed in section V (B) ) is also tested. It is requested to repeat the workflow with WfID 114. It first collects the resource configuration from the database and provisions resources on the Cloud infrastructure. The name of re-provisioned resource(s) for the repeated workflow has a postfix '-rep.novalocal' e.g.mynova-rep.novalocal as shown in Table II. It was named mynova.novalocal in original workflow execution as shown in Table I. From Table II, one can assess that similar resources have been re-provisioned to repeat the workflow execution because the RAM, Hard disk, CPUs and image configurations are similar to the resources used for workflow with WfID 114 (as shown in Table I). This preliminary evaluation confirms that the similar resources on the Cloud can be re-provisioned with the Cloud-aware provenance information collected using the proposed approach (discussed previously in section IV). Table II shows two repeated instances of original workflow 114.

VI. RELATED WORK

Significant research [16, 17, 18] has been carried out in workflow provenance for Grid-based workflow management systems. Chimera [16] is designed to manage the derivation and analysis of data for high-energy physics (GriPhyN) [30] and astronomy (SDSS) [31] communities, which are data-intensive. It captures process information along with the parameters used and the data used as input and the produced data. It stores this provenance information in Chimera schema, which is based on a relational database. Although the schema allows storing the physical location of a machine, it does not support the state of hardware and software environment in which a transformation executes. Vistrails [17] is a workflow and provenance management system that provides support for scientific data exploration and visualization. It not only captures the execution log of a workflow but also the changes a user makes to refine their workflow. However, it does not support the infrastructure layer information of the Cloud infrastructure. Similarly, Pegasus/Wings [34] records evolution of a workflow and records all edit operations. However, this work is based on the workflow execution provenance, rather than the provenance of a workflow itself (e.g. design changes).

There have been a few research studies [19, 20, 22] performed to capture provenance in the Cloud. However, they lack the support for workflow repeatability. Some of the work towards provenance in the Cloud is directed to the file system [27, 20, 21] or hypervisor level [28] this provenance information, is not relatable to this work. The PRECIP [10] project is the closest to this proposed work. However, PRECIP provides an API to provision and execute workflows. It does not capture Cloud infrastructure information and establish mappings between workflow jobs and the Cloud infrastructure resources. Missier et al. [33] proposed an approach that compares the provenance traces of two given workflows to verify the experiment reproducibility and identifies the divergence in provenance through provenance graph analysis. It also compares the outputs using different tools such as GNU diff[4] for text files, and other eScience tools for complex datasets. However, our initial prototype uses hash values of the produced outputs, which are text at the moment and they do not consider Cloud information.

There have been a few recent projects [36, 37] and research studies [38] on collecting provenance and using it to reproduce an experiment. Santana-Parez et. al [38, 39] proposes the use of semantics to improve reproducibility of workflows in the Cloud. In doing so, it uses ontologies to extract information about the computational environment from the annotations provided by a user. This information is then used to recreate (install or configure) that environment to reproduce a workflow and its execution. On the contrary, our approach is not relying on annotations rather it directly interacts with the Cloud middleware at runtime to acquire resource configuration information and then establishes mapping between workflow jobs and Cloud resources. Similarly, a research study [40] has proposed to conserve the VM that is used to execute a job and then reuse the same VM while re-executing the same job. One may argue that it would be easier to keep and share VM images with the community research through a common repository, however the high storage demand of VM images remains a challenging problem [41]. Moreover, keeping VM image information is not enough for re-provisioning a resource on the Cloud. The ReproZip software [36] uses provenance and system call traces to provide reproducibility and portability. It can capture and organize files/libraries used by a job. All this information is collected in a configuration file, and all the files are zipped together for portability and reproducibility purposes. Since this approach is useful at individual job level, this does not work for an entire workflow, which is the focus of this research paper. Moreover, this approach does not consider the hardware configuration of the underlined execution machine. Similarly, a Linux-based tool, CARE [37], is designed to reproduce a job execution. It builds an archive that contains selected executable/binaries and files accessed by a given job during an observation run.

---

[4] gnu.org/software/diffutils [Last access 05-08-2014]

## VII. CONCLUSION AND FUTURE WORK

In this paper, the motivation and the issues related to workflow repeatability due to workflow execution on the Cloud infrastructure have been identified. The dynamic nature of the Cloud makes provenance capturing of workflow(s) and their underlying execution environment(s) and their repeatability a difficult challenge. A proposed architecture has been presented that can augment the existing workflow provenance with the information of the Cloud infrastructure. Combining these two can assist in re-provisioning the similar execution environment to repeat a workflow execution. The Cloud infrastructure information collection mechanism has been presented in this paper in section IV (A). This mechanism iterates over the workflow jobs and establishes mappings with the resource information available on the Cloud. This job to Cloud resource mapping can then be used to repeat a workflow. The process of repeating a workflow execution with the proposed approach has been discussed in section IV (B). A prototype was developed for evaluation and the results showed that the proposed approach creates a similar execution infrastructure i.e. same resource configuration on the Cloud using the Cloud-aware provenance information (as discussed in section IV) (see Fig. 4) to repeat a workflow execution. In future work, the proposed approach will be extended and a detailed evaluation of the proposed approach will be conducted. Different performance matrices such as the impact of the proposed approach on workflow execution time and total resource provision time will also be measured. In future, more emphasis will be given to the mechanisms to incorporate the workflow output comparison along with infrastructure comparison to verify workflow repeatability.

ACKNOWLEDGMENT

This research work has been funded by a European Union FP-7 project, N4U – neuGrid4Users. This project aims to assist the neuro-scientific community in analysing brain scans using workflows and distributed infrastructure (Grid and Cloud) to identify biomarkers that can help in diagnosing the Alzheimer disease. Besides this, the support provided by OSDC by offering a free Cloud infrastructure of 20 cores is highly appreciated. Such public offerings can really benefit research and researchers who are short of such resources.